# Present and future searches with e⁺e⁻ colliders for the neutral Higgs bosons of the Minimal Supersymmetric Standard Model – the complete 1-loop analysis

Janusz Rosiek*

Institut de Física Corpuscular, Universitat de Valencia
Dr. Moliner 50, 46100 Burjassot, Valencia, Spain

André Sopczak

PPE Division, CERN
CH-1211 Geneva 23, Switzerland

**Abstract**

New mass regions unexcluded by direct searches are revealed by an analysis of experimental results from LEP1 using full 1-loop diagrammatic calculations of radiative corrections in the MSSM. Simulations of experimental signal efficiencies and background rejection factors, and full 1-loop calculations are combined to study the sensitivity for neutral Higgs bosons at LEP2 and the NLC. Compared with previous studies based on an Effective Potential Approach, we identify mass regions where the discovery potential depends on the MSSM parameters other than the top and stop masses. We propose our method of interpretation to be adopted by the four LEP experiments for better precision. The possibility of the experimental distinction between the Higgs boson in the MSM and the lightest Higgs boson in the MSSM is discussed for the NLC.

# 1. Introduction

The search for Higgs particles is one of the most challenging problems of experimental particle physics. The experimental evidence for Higgs boson(s) is crucial to understand the mechanisms of the SU(2) × U(1) symmetry breaking and the mass generation in gauge theories. At present, no Higgs bosons have been found and only lower bounds on Higgs boson masses are established. The most stringent bounds come from the LEP1 collider running at $\sqrt{s} \approx m_Z$. In the future, e$^+$e$^-$ collider experiments will continue to play a very important role in the hunt for Higgs bosons because of their low background and clean signatures. Particular attention is given to the search for the Higgs bosons with properties predicted by the Minimal Supersymmetric Standard Model (MSSM). The MSSM has many promising features and it is the most discussed extension of the Minimal Standard Model (MSM). The Higgs sector of the MSSM contains five physical Higgs bosons, one neutral CP-odd scalar, A$^0$, two neutral CP-even scalars, h$^0$ and H$^0$, and two charged scalars, H$^\pm$. At least one of them, the lighter CP-even scalar h$^0$, is expected to be relatively light ($m_h \leq 130 - 140$ GeV for $m_t \approx 175$ GeV) and its discovery could be the first signal for supersymmetry. Consequently, we focus on the interpretation of searches for the neutral Higgs bosons within the MSSM and on their perspectives at future experiments.

Numerous results of searches for Higgs bosons in the framework of the MSSM are reported from the LEP1 experiments Aleph, Delphi, L3, and Opal [1] (for an overview see [2]). In the next phase of the LEP programme, LEP2, $\sqrt{s} > 2m_W$ is planned to be reached in 1996 [3]. The physics potential of the e$^+$e$^-$ Next Linear Collider, NLC, running at much higher energies, $\sqrt{s} \geq 500$ GeV, is intensively discussed [4].

A realistic analysis of the phenomenology of the MSSM Higgs sector has to include radiative corrections [5-11]. These corrections can strongly modify the Higgs boson masses and couplings. At the tree-level, the Higgs sector of the MSSM can be effectively parameterized in terms of two free variables, for example, the masses of two Higgs bosons. After including the 1-loop corrections, Higgs boson masses and couplings to depend also on the top quark mass and the additional unknown parameters of the MSSM. Throughout the paper, these unknown parameters are referred to as SUSY parameters. Several approaches have been developed to compute radiative corrections to the tree-level approximation:

- the Effective Potential Approach (EPA) [6],

- the Renormalization Group Equations (RGE) approach [7, 8], and

- the Full 1-loop Diagrammatic Calculations (FDC) in the on-shell renormalization scheme [9, 10].

Previous interpretations in the MSSM of searches for Higgs bosons, which are performed by the LEP experiments [1], are based on common assumptions:

1. Radiative corrections to the MSSM Higgs boson masses, production and decay rates are considered in the EPA approximation, in its most simplified version. Only the leading part of the contribution arising from the top quark and from its supersymmetric scalar partner, stop, is considered.

2. The MSSM parameter space is strongly constrained by neglecting the dependence on most of the SUSY particle masses and couplings. In the most simplified version of the EPA, only one free SUSY parameter, $m_{\tilde{t}}$ is considered.

3. In the EPA interpretations it is assumed that $m_{\tilde{t}} > m_t$, which is experimentally unjustified.

In two aspects, we improve the interpretations of existing experimental results and perspectives of future searches for the neutral MSSM Higgs bosons. First, we interpret existing LEP1 data and simulations of detector efficiencies expected for LEP2 and the NLC using more accurate theoretical calculations of the production and decay rates of $h^0$, $H^0$ and $A^0$ in the FDC approach, and then we compare the results with the EPA predictions. Second, we study the effect of using a larger set of parameters describing the MSSM, taking into account the existing experimental constraints on the SUSY parameters.

In order to compute the theoretical predictions for the production cross sections and the decay branching ratios, the full 1-loop diagrammatic calculations in the on-shell renormalization scheme have been used (a detailed description of the FDC method can be found in [9]). This approach takes into account the virtual effects of all possible MSSM particles and includes contributions which have yet been neglected in the EPA, such as gauge sector contributions, momentum-dependent effects in 2- and 3-point Green's functions, and genuine 1-loop corrections to 3-point functions. The most significant difference between our approach and the EPA comes from the chargino and neutralino sector contributions.

Experimental searches should rely as little as possible on theoretical assumptions. Therefore, we consider a much less constrained parameter space compared to previous studies. Using the FDC we are able to explore the dependence of the masses and couplings of the Higgs bosons on *all* soft breaking parameters in the MSSM Lagrangian. Some relations and simplifications are applied in order to decrease the number of free parameters, after checking the results in the Higgs sector to be not sensitive to those assumptions.

We discuss three main aspects of the neutral MSSM Higgs boson searches: implications of the existing experimental data from LEP1, perspectives of the searches at LEP2, and at the NLC. Results are presented as:

- Mass regions excluded by the LEP1 data. The results presented here are based on the measurements from the L3 collaboration. The published data set corresponding to about 408 000 hadronic $Z^0$ decays has been used [12].

- The Higgs boson discovery potential of the LEP2 collider for center-of-mass energies 175 GeV, 190 GeV and 210 GeV. Our analysis is based on results of simulations of the expected sensitivity of searches for Higgs bosons [13].

- An analysis of the discovery potential of the NLC running at $\sqrt{s} = 500$ GeV. The possibilities of distinguishing the lightest Higgs boson in the MSSM from the Higgs boson of the MSM are discussed.

One should note that it is not feasible to derive a precise combined LEP1 limit on the Higgs mass owing to the different methods used by the four LEP experiments which lead to different detection efficiencies and number of data and background events. A simple combined confidence level calculation, based on the number of expected Higgs bosons, data and background events, tends to overestimate the sensitivity. This is due to the fact that the experimental detection sensitivities are tuned by each experiment on the basis of their own data set using the appropriate selection cuts. An increase of the amount of data (corresponding to combined LEP statistics) would result in tighter selection cuts which are then derived from the larger expected background sample. Therefore, the signal detection efficiency decreases with increasing statistics which affects the overall increasing signal sensitivity.

## 2. Parameter Space of the MSSM

The most general version of the MSSM Lagrangian contains a large number of free parameters. Most of the SUSY parameters have small impact on the Higgs sector. Using numerical simulations, we identified which parameters are important for the Higgs boson phenomenology. These parameters have been varied independently:

- $(m_\mathrm{h}, m_\mathrm{A})$ or $(m_\mathrm{H}, m_\mathrm{A})$ – the investigated Higgs boson mass combinations.

- $m_\mathrm{sq}$ – the common mass parameter for all squarks. The assumption of the same mass parameters for the three squark generations has a small effect. Results depend mostly on the stop mass parameter and only weakly on the masses of other sfermions.

- $m_\mathrm{g}$ – the gaugino mass. We assumed the commonly used GUT relation for the SU(2) and U(1) gaugino masses: $m_\mathrm{U(1)} = \frac{5}{3}\tan^2\theta_W m_\mathrm{SU(2)}$, $m_\mathrm{SU(2)} = m_\mathrm{g}$. This assumption also has a little impact on results.

- $\mu$ – the mixing parameter of the Higgs doublets in the superpotential.

- $A$ – the mixing parameter in the sfermion sector. As for $m_\mathrm{sq}$ only one universal mixing parameter is considered for all squark generations. The mixing is proportional to $A m_\mathrm{sq}$.

Recently, the CDF collaboration reported direct evidence for the top quark, compatible with a mass of $174 \pm 10^{+13}_{-12}$ GeV [14]. Throughout our paper, the top quark mass is fixed to $m_\mathrm{t} = 175$ GeV. In order to study the effect of the variation of the SUSY parameters described above we scan them in the ranges given in Table 1. The parameters shown in

| Parameter | $m_\mathrm{sq}$ (GeV) | $m_\mathrm{g}$ (GeV) | $\mu$ (GeV) | $A$ |
|---|---|---|---|---|
| Range | 200 — 1000 | 200 — 1000 | −500 — 500 | −1 — +1 |

Table 1: Ranges of SUSY parameters used for independent variation in the study of the MSSM neutral Higgs boson searches.

Table 1 are the input parameters for the calculations of the physical sfermion, chargino, and neutralino masses. Some parameter combinations can be unphysical (e.g. negative squark masses) or experimentally excluded. Such cases are removed by imposing the following constraints: stop and chargino are required to be heavier than $m_Z/2$, and the neutralino to be heavy (or weakly coupled) in agreement with the bound on contributions to the $Z^0$ width beyond the MSM: $\Delta\Gamma_Z^{\max} < 31$ MeV [15].

An additional constraint is applied on $\tan\beta$, defined as the ratio of the vacuum expectation values of the Higgs doublets. Although at the 1-loop level $\tan\beta$ is renormalization scheme dependent, this dependence is rather weak [9]. Tree-level experimental bounds on $\tan\beta$ are assumed to hold approximately, and its value is constrained to the range of $0.5 \leq \tan\beta \leq 50$. The lower bound is based on [16]. The variation of the upper bound has no significant effect on the results. In our approach $\tan\beta$ is a function of $m_h$, $m_A$ and the SUSY parameters listed in Table 1. The lower bound on $\tan\beta$ affects the theoretically allowed regions in the $(m_h,m_A)$ and $(m_H,m_A)$ planes. The change of $\tan\beta > 1$ to $\tan\beta > 0.5$ extends the theoretically allowed region for $h^0$ mass of about $\Delta m_h = 20$ GeV for $m_h > m_A$.

This analysis required a large scale numerical simulation due to the complexity of the FDC method. The multidimensional sampling over the parameters described above required several thousand hours of CPU time.

## 3. Uncertainties in the Predictions

Several possible sources of uncertainties exist in the interpretation of the experimental results in the framework of the MSSM. Two of them have been addressed by taking into account a larger set of the MSSM parameters, and by using FDC as the most accurate method to calculate the 1-loop radiative corrections. We find that for fixed values of $m_A$ and $\tan\beta$ and for an equivalent set of SUSY parameters, the FDC gives typically $m_h$ values reduced by 5–7 GeV in comparison to EPA. The cross sections $e^+e^- \to Z^0h^0, A^0h^0$ differ typically by 15-30%, the difference can be much larger for some sets of model parameters.

As an example, Fig. 1 illustrates the dependence on the SUSY parameters of the upper limits on $m_h$ as a function of $m_A$. For comparison, the EPA predictions are plotted. Thin and thick lines illustrate these limits for $m_t = 130$ GeV and 180 GeV, respectively. Cross-marked lines are plotted assuming all SUSY particles to be heavy ($\approx 1$ TeV) and no mixing in the sfermion sector ($A = 0$). The values of these parameters are listed in Table 2. Dotted lines show the case of light squarks ($\approx 200$ GeV) (the contribution of sleptons is negligible). The top quark and stop masses have the largest impact on the upper $m_h$ bound. However, from the experimental point of view, the dependence on gaugino masses and on mixing in the squark sector can also be very important: for instance changing $m_g$ from 1 TeV (cross-marked) to 200 GeV (dashed) shifts $m_h$ by $\mathcal{O}(5\text{ GeV})$. The shift can be positive or negative, depending on the other parameters. Solid lines show the absolute upper bound on $m_h$ obtained by the independent variation of all parameters in the ranges defined in Table 1. Circle-marked lines show the upper bound on $m_\mathrm{h}$ obtained in the EPA for the same set of SUSY parameters (Table 2) as the cross-marked lines. The EPA upper bound is higher by about 5 GeV. A scan over the full set of SUSY parameters overcompensates the reduction and leads to a SUSY parameter-independent upper bound on $m_\mathrm{h}$ higher by $\Delta m_\mathrm{h} = 7$ GeV. This difference is particularly important in the study of the kinematically accessible mass region at LEP2.

| Parameter | $m_{sq}$ (GeV) | $m_g$ (GeV) | $\mu$ (GeV) | $A$ |
|---|---|---|---|---|
| Value | 1000 | 1000 | 100 | 0 |

Table 2: Fixed heavy SUSY parameters used for comparison with the EPA results

The other possible sources of uncertainties are connected with the increase of the range of the variation over the MSSM parameter space and with the higher order radiative corrections. We have checked that the change of the limits shown in Table 1 has only a small effect on the results discussed in next sections. A decrease of the lower limit for the sfermion and gaugino masses causes in most cases at least one SUSY particle to be light and observable. Such parameter combinations are rejected. An increase of the upper limits of the sfermion and gaugino masses influences the upper limit on the $h^0$ mass. This effect is rather weak as it depends only logarithmically on the squark masses. The increase of the range of the sfermion mixing parameter $A$ and taking into account more general mixing terms can be more significant. Such possibilities are, however, unlikely both from the theoretical and experimental point of view. They are difficult to be generated in Grand Unified Theories and they can have significant effects in many low-energy processes. Large $A$ values lead to large splitting of the sfermion masses and usually to the existence of a light sfermion, which should be directly visible in existing experiments. Non-diagonal soft-breaking couplings are the source of sizeable flavor-changing neutral currents (and, if complex, CP breaking effects [17]), which are ruled out by existing data. Therefore, the choice of the bounds shown in Table 1 is well motivated and results are found to be stable against small variations.

Finally, some estimates are given of the 2-loop corrections for the MSSM Higgs boson masses [18]. The published results are at least partially inconsistent with each other. A recent detailed study [19] shows that with the correct definition of the physical top quark mass as the pole of the propagator (consistent with our FDC method) 2-loop corrections to the $h^0$ mass are negative and small. No significant effects on our results are expected.

## 4. Excluded Mass Regions at LEP1

In order to derive precise bounds on $h^0$ and $A^0$ masses, the limits on the Higgs boson production rates given in [12] have been used. In the mass plane ($m_\mathrm{h}$, $m_\mathrm{A}$), each point with a step size of 1 GeV up to Higgs boson masses of 120 GeV has been analyzed separately. For each mass combination, the production cross sections of the reactions[1] $e^+e^- \to h^0 Z^{0\star} \to h^0 \overline{f} f$, $e^+e^- \to h^0 A^0$, and the branching ratios for $h^0$ and $A^0$ decays have been computed as a function of the parameters described in Sec. 2. Then, the number of expected Higgs boson events for each investigated final state and each mass bin has been calculated. The following channels are taken into account:

1) $h^0$ production in bremsstrahlung processes:

---

[1] The possibility of Higgs boson production via bremsstrahlung off b-quark $e^+e^- \to b\overline{b} \to b\overline{b}h^0$ is not discussed. This channel could be significant for large values of $\tan\beta$ [20]. Also the fusion of $W^+W^-$ is not considered as it is negligible for LEP1 and LEP2 energies. On the contrary, this reaction could become significant at the NLC [21].

$$e^+e^- \rightarrow h^0 Z^{0\star} \rightarrow h^0 e^+e^-, h^0\mu^+\mu^-, h^0\bar{\nu}\nu$$

2) $h^0 A^0$ pair-production processes:

$$e^+e^- \rightarrow h^0 A^0 \rightarrow \tau^+\tau^-\tau^+\tau^-, \tau^+\tau^-\bar{b}b, b\bar{b}b\bar{b}.$$

For $m_h > 2m_A$:

$$e^+e^- \rightarrow h^0 A^0 \rightarrow A^0 A^0 A^0 \rightarrow \bar{b}b\bar{b}b\bar{b}b.$$

In addition, a combined LEP1 limit on non-standard $Z^0$ decays has been applied: $\Delta\Gamma_Z^{\max} < 31$ MeV at 95% CL [15].

A given $(m_h, m_A)$ combination is excluded if for *all* SUSY parameter sets (from the ranges defined in Table 1 and for fixed $m_t = 175$ GeV) the expected number of events in at least one of the channels is excluded at 95% CL.

Figure 2 shows regions in the $(m_h, m_A)$ plane which are excluded by the individual channels listed above. A comparison of the excluded regions of Fig. 2 with the combined excluded region of Fig. 3 shows that the sum of the partial exclusion regions is smaller than the combined one. This is due to the scanning over the SUSY parameters. For fixed $(m_h, m_A)$, one can find the parameter combinations for which the cross section for a given channel is particularly low. It is unlikely that the cross sections are very low in all channels simultaneously, owing to the well-known complementarity of the cross sections of $e^+e^- \rightarrow h^0 Z^0$ and $e^+e^- \rightarrow h^0 A^0$ reactions. At the tree-level, production rates are proportional to $\sin^2(\alpha - \beta)$ and $\cos^2(\alpha - \beta)$. This complementarity holds approximately even after the inclusion of non-leading vertex corrections. The excluded regions in the $(m_A, \tan\beta)$ plane are presented for each channel separately in Fig. 4.

Figures 3 and 5 show the regions in the $(m_h, m_A)$ and $(m_A, \tan\beta)$ planes that can be excluded by the simultaneous analysis of all channels. Three regions are distinguished:

i) Excluded regions after performing a full scan over the SUSY parameter space and using the FDC method in cross section and branching ratio calculations;

ii) as above, but varying only $m_{sq}$ and assuming that the other SUSY parameters are constrained to the values shown in Table 2. This is done for comparison with the EPA approximation;

iii) excluded regions with radiative corrections calculated in the simplified EPA ("epsilon approximation") [12], where only the leading corrections from the top and stop loops are taken into account. In this case results depend on $m_{sq}$ only. The range $175 \text{ GeV} \leq m_{sq} \leq 1000 \text{ GeV}$ is used.

Figure 3 reveals an interesting result for the excluded regions in the $(m_h, m_A)$ plane. The full scan over the SUSY parameter space (thick solid line) gives, in comparison with the epsilon approximation (dotted line, iii), a substantial additional triangle-shape unexcluded mass range for $45 \text{ GeV} < m_A < 80 \text{ GeV}$ and $25 \text{ GeV} < m_h < 50 \text{ GeV}$, which is marked with a bold solid line. The existence of this region can be understood in the following way: in the range $m_h + m_A < m_Z$ the reaction $e^+e^- \rightarrow h^0 A^0$ is allowed kinematically and both main discovery channels $e^+e^- \rightarrow h^0 Z^{0\star}$, and $e^+e^- \rightarrow h^0 A^0$ contribute. If radiative corrections reduce the cross section of one of them below the experimental sensitivity, the complementary cross section will be large enough to exclude this mass combination. The unexcluded triangle begins just above the $m_h + m_A = m_Z$ limit. In this range the bremsstrahlung cross section $e^+e^- \to h^0 Z^{0\star}$ can be small for some SUSY parameters. We identify points where it is suppressed by a factor of 25 compared with the MSM prediction for $e^+e^- \to H^0_{MSM} Z^{0\star}$. The complementary process is already forbidden kinematically, thus no signal can be observed.

An example for an unexcluded mass point is given in Table 3. The cross sections for Higgs boson bremsstrahlung and pair-production obtained in FDC are listed for a chosen set of SUSY parameters. The simple version of the EPA used by the LEP experiments does not allow $\sqrt{m_{\tilde{t}1} m_{\tilde{t}2}} < m_t$, therefore, no corresponding cross section exists for the given FDC example. Low unexcluded $m_h$ values are obtained for low physical stop masses of the order of $\mathcal{O}(50 - 200 \text{ GeV})$ and large mixing in the sfermion sector ($A = \pm 1$, large $\mu$). In such cases the splitting between the left and right stop masses is large. The other SUSY parameters have smaller influence on the shape of the unexcluded region. With increasing $m_A$, the cross section for the $e^+e^- \to h^0 Z^{0\star}$ reaction becomes less sensitive to the SUSY parameters and similar to the $e^+e^- \to H^0_{MSM} Z^{0\star}$ cross section (calculated at $m_h = m_{H^0_{MSM}}$) because of the known decoupling effect [8, 23]. The difference between cross sections calculated in the MSSM and MSM decreases as $1/m_A^4$. Above $m_A \approx 100$ GeV the bremsstrahlung production of $h^0$ is sufficient to establish, independent of the SUSY parameters, the Higgs mass bound of 55 GeV. Even in this range of $m_A$, for special SUSY parameter combinations (outside the values defined in Table 1), a light $h^0$ can escape detection for very large squark mixing. Such combinations, however, are unlikely from the theoretical and experimental point of view, as discussed in the Sec. 3.

| $m_h$ | $m_A$ | $m_t$ | $m_{sq}$ | $m_g$ | $\mu$ | $A$ | $\tan\beta$ | $m_{\tilde{t}1}$ | $m_{\tilde{t}2}$ | $\sigma_{hA}$ | $\sigma_{h\nu\bar{\nu}}$ |
|---|---|---|---|---|---|---|---|---|---|---|---|
| 32 | 64 | 175 | 200 | 200 | $-500$ | $-1$ | 3.2 | 81 | 362 | 0.0 | 0.25 |

Table 3: Example of an unexcluded mass point. Cross sections for Higgs boson bremsstrahlung and pair-production for a chosen set of SUSY parameters. All mass parameters are given in GeV, and cross sections in pb.

Figure 3 shows that the regions obtained in approaches (ii) (thin solid line) and (iii) (dotted line) are similar. The excluded area in (ii) is only slightly larger than in (iii). The few GeV distance between the lines reflects the difference in the $m_h$ values calculated in the EPA and the FDC. This shows that the EPA result can be approximately recovered for a specific set of SUSY parameters given in Table 2.

Figure 5 shows the results in the $(m_A, \tan\beta)$ plane. No significant differences in the shape of the excluded regions obtained in the EPA (dotted line) and the full scan FDC (thick solid line) is visible. This parameterization is less suited to investigate the experimental signal. We emphasize that the $(m_h, m_A)$ variables are more useful and natural from the experimental point of view and in this parameterization significant differences between both approaches become obvious.

The size of the excluded regions presented in Figs. 3 and 5 are rather insensitive to the choice of a lower bound on $\tan\beta$. The $\tan\beta$ values from the range 0.5–1 are projected to $m_h$ values ranging from 60 GeV to 80 GeV, which is almost entirely above the reach of LEP1. Hence, the assumption $\tan\beta \geq 0.5$ affects only the theoretical upper bound of the $h^0$ mass range allowed in the MSSM.

## 5. LEP2 Discovery Potential

Four production reactions relevant for LEP2 $e^+e^- \to h^0 Z^0, h^0 A^0$ and $e^+e^- \to H^0 Z^0, H^0 A^0$ have been investigated based on sensitivities given in [13]. As in the case of LEP1, the effect of the variation of the SUSY parameters has been studied for the fixed values of $m_h$ and $m_A$. Each point in the ($m_h$,$m_A$) plane has been analyzed separately with a step size of 5 GeV. For each fixed mass combination (or fixed $m_A$ and $\tan\beta$), the production cross sections of all the four reactions have been calculated for $\sqrt{s} = 175$, 190 and 210 GeV as a function of the parameters listed in the Sec. 2. When the first signal is visible, $h^0$ and $H^0$ are indistinguishable owing to low production rates and similar signatures. As a consequence, a given point in the parameter space is accessible at LEP2 if at least one of the cross sections $\sigma_{hZ}$, $\sigma_{hA}$, $\sigma_{HZ}$ or $\sigma_{HA}$ is larger than the expected experimental sensitivity. Such a mass point is called a sensitivity point. For Higgs boson bremsstrahlung and pair-production, similar detection sensitivities can be expected. Their precise values depend on the achievable signal efficiency and the reduction of the background. A linear interpolation has been used to obtain the sensitivity for mass combinations between simulated mass points. Four regions are distinguished in the ($m_h$,$m_A$) and ($m_A$,$\tan\beta$) planes:

(A) The sensitivity region where, by direct searches, a Higgs signal cannot escape detection, for any choice of the SUSY parameters from the ranges given in Table 1 and for fixed top quark mass of 175 GeV.

(B) The region where the perspectives of direct searches depend on the SUSY parameters. This means that searches can have sensitivity or not, depending on the specific choice of these parameters.

(C) The non-sensitivity region where no signal can be found independent of the choice of the SUSY parameters.

(D) The theoretically disallowed region in the ($m_h$,$m_A$) parameterization where ($m_h$,$m_A$) combinations are not allowed in the MSSM for any choice of SUSY parameters and requiring $\tan\beta \geq 0.5$.

The mass regions where at least one CP-even Higgs boson, $h^0$ or $H^0$, can be discovered at LEP2 for $\sqrt{s} = 175$, 190 and 210 GeV are shown in the ($m_h$,$m_A$) and ($m_A$,$\tan\beta$) planes in Figures 6 and 7, respectively.

The effects of increasing center-of-mass energy can be clearly seen in Fig. 6. A substantial region (B) reflects an uncertainty in the discovery potential connected with the variation over the SUSY parameters. The border between the regions (B) and (C) is largely set by the kinematical bound for the bremsstrahlung process $e^+e^- \to h^0 Z^0$ and depends strongly on the available center-of-mass energy. The upper bound of region (C) depends mainly on the top quark mass. For $m_A > 100$ GeV, region (B) forms a band about 10 GeV wide, tangent to the kinematical bound. The second part of region (B) lies in the intermediate region $m_h \approx m_A \leq 120$ GeV. For some combinations of the SUSY parameters even $h^0$ as light as about 60 GeV can escape detection. Cross sections in this region are not much below the detection sensitivities and a fine-tuned experimental analysis can probably cover the low-$m_h$ part of region (B).

Figure 7 shows the same results in the ($m_A$,tan $\beta$) plane. The regions (A) are similar for $\sqrt{s}$ = 175 and 190 GeV. For $\sqrt{s}$ = 210 GeV the bound of region (A) shifts by about 15 GeV to about 70 GeV. For $\sqrt{s}$ = 175 GeV a large region (C) reflects the fact that there is no possibility to discover a MSSM scalar for $m_A > 60$ GeV and tan $\beta > 5$. Region (C) shrinks to small isolated area for $\sqrt{s}$ = 190 GeV and it is entirely replaced by region (B) for $\sqrt{s}$ = 210 GeV. This shows that for the higher center-of-mass energies the bremsstrahlung reaction $e^+e^- \to h^0Z^0$ can be observed even for large $m_A$ values, with the exception of some choices of the SUSY parameters. For $\sqrt{s}$ = 210 GeV a small additional region (A) appears around the $m_A = 100 - 120$ GeV and tan $\beta > 40$ due to $H^0$ contributions. Both regions (A) expand with further increase of the center-of-mass energy. Comparing the presentation ($m_h$,$m_A$) and ($m_A$,tan $\beta$) one should note that several points of ($m_h$,$m_A$) can correspond to one point in ($m_A$,tan $\beta$). Therefore, region (C) does not exist in the ($m_A$,tan $\beta$) presentation, while it exists in the ($m_h$,$m_A$) parameterization.

The theoretically allowed mass regions depend strongly on the lower tan $\beta$ bound. The assumption tan $\beta > 0.5$ gives in comparison with tan $\beta > 1$ an additional range 60 GeV < $m_h$ < 80 GeV for $m_A < m_Z$. However, this region is in the range of LEP2. For $m_A > m_Z$, the upper $m_h$ bound corresponds to large tan $\beta$ values.

Concluding, the mass ranges accessible for LEP2 depend strongly on the experimentally achievable center-of-mass energy. If the top quark mass is close to or higher than the central value of CDF measurements [14], even for $\sqrt{s}$ = 210 GeV and $\mathcal{L}$ = 500 pb$^{-1}$ LEP2 cannot perform a decisive test of the MSSM. Most of the allowed ($m_h$,$m_A$) plane is covered, but some mass regions remain out of reach also for this machine configuration.

## 6. Discovery potential of the NLC

In the analysis of the physics potential of the NLC we use a center-of-mass energy of $\sqrt{s}$ = 500 GeV and an estimated sensitivity of 10 fb for all $e^+e^- \to h^0Z^0$, $H^0Z^0$, $h^0A^0$, and $H^0A^0$ channels [4]. Assuming a total luminosity of the NLC of $\mathcal{L}$ = 30 fb$^{-1}$, this sensitivity corresponds approximately to a discovery of a signal if more than 300 events are produced (before selection cuts are applied). Under these assumptions the NLC can cover entirely the MSSM parameter space and at least one Higgs boson must be found or the MSSM is ruled out. This conclusion holds for a simultaneous search for the pair production reactions $e^+e^- \to h^0A^0, H^0A^0$ and for the bremsstrahlung reactions $e^+e^- \to h^0Z^0, H^0Z^0$. The reaction $e^+e^- \to h^0Z^0$ alone is not sufficient because the cross section for this process is too low to be discovered for some parameter choices. At the NLC good chances exist to find more than one Higgs boson if its mass is not too large. Figure 8 illustrates the perspectives of finding the heavier CP-even Higgs bosons $H^0$. Regions (A)–(D) are defined as in Sec. 5. Some fraction of the parameter space for $m_A \leq 100$ GeV is covered by searches for $H^0$ bremsstrahlung only. The remaining region can be covered by searches for $H^0A^0$ pair-production up to about $m_A + m_H \leq 400$ GeV.

Figure 9 shows regions in the ($m_h$,$m_A$) plane where more than one MSSM Higgs boson could be found. This is of particular interest, since the discovery of more than one Higgs boson most clearly distinguishes the MSSM from the MSM. The important conclusion resulting from the analysis of the production rates of the neutral MSSM Higgs bosons at the NLC is that either $h^0$ alone, or all three neutral MSSM scalars $h^0$, $H^0$ and $A^0$ could be found simultaneously. This is due to the complementarity of the couplings $Z^0Z^0h^0$, $Z^0H^0A^0$ and $Z^0Z^0H^0$, $Z^0h^0A^0$. For $m_A > 100$ GeV, the decoupling effect becomes important [8, 23] and as a result the cross section for the $e^+e^- \to Z^0h^0$ process and the $h^0$ decay branching ratios are close to the MSM predictions. In the same $m_A$ range, the $Z^0H^0A^0$ coupling is strong and $Z^0Z^0H^0$ is weak, thus $H^0$ can only be produced in association with $A^0$ in the $e^+e^- \to H^0A^0$ reaction. For smaller $A^0$ masses and some SUSY parameter choices, $h^0$ bremsstrahlung cannot be observed. In this case the $Z^0Z^0H^0$ and $Z^0h^0A^0$ couplings are large and both process are kinematically allowed. Again all three neutral MSSM scalars could be detected. This conclusion also holds after taking into account the $W^+W^-$ fusion: $e^+e^- \to W^+W^-\nu\bar{\nu} \to \nu\bar{\nu}h^0(H^0)$, since the $W^+W^-h^0(H^0)$ and $Z^0Z^0h^0(H^0)$ couplings are proportional.

Figure 9, region (A), shows that all three scalars could be observed up to $m_A \approx 200$ GeV. For $m_h \approx 90$ GeV and $m_A = 180$ GeV a small region (B) exists, where the perspectives of the simultaneous $h^0$, $H^0$ and $A^0$ discovery depend on the SUSY parameters. For larger $m_A$ only $h^0$ can be found, region (C).

## 7. Distinction between MSSM and MSM Higgs Bosons

Various possibilities exist to identify a Higgs boson unambiguously as a non-minimal one after a first discovery. The observation of two (or more) Higgs scalars would be very interesting, since it immediately excludes the MSM and gives support to multi-doublet Higgs models, with the MSSM as a possible example[2]. Additional possibilities of distinguishing the lightest Higgs boson in the MSSM from the MSM one are important in the regions of the parameter space where only the $h^0$ could be observed (large $m_A$). In those regions the distinction between the MSSM and MSM could eventually be made on the basis of measuring the coupling of an observed scalar to the $Z^0$. However, the cross sections for the processes $e^+e^- \to Z^0h^0$ and $e^+e^- \to Z^0H^0_{MSM}$ are very similar for large $m_A$. For $m_A > 100$ GeV the difference is typically smaller than several percent [23]. Even for $m_A < 100$ GeV, the cross section difference can be small, depending on the SUSY parameters. Owing to the small number of expected Higgs events, such difference is probably not sufficient to assess the origin of the observed scalar. Figure 10 illustrates where the difference $|(\sigma_{MSSM} - \sigma_{MSM})/\sigma_{MSM}|$ is larger than 20%. The definitions of the regions are the following:

(A) Region where the difference of the cross section for Higgs boson bremsstrahlung is always larger than 20%, independent of the choice of the SUSY parameters (varied in the ranges defined in Table 1).

(B) Region where the cross section difference is larger or smaller than 20%, depending on the SUSY parameters.

---

[2]In this paper, only the neutral MSSM scalars are discussed. For a large part of the MSSM parameter space charged Higgs bosons could be discovered. Their discovery would also give rise to unambiguous evidence of physics beyond the MSM. Simulations in the EPA approximation [22] show that experiments at NLC can detect charged Higgs bosons with masses up to about 210 GeV, what corresponds to $m_A \approx 200$ GeV.

(C) Region where the cross section difference is smaller than 20% for any choice of the SUSY parameters.

(D) Region disallowed in the MSSM.

Another method of identifying the origin of a scalar is based on measuring its decay branching ratios. The dominant channel of $h^0$ decays under consideration is $h^0 \to b\bar{b}$. Supersymmetric decay modes could allow an effective distinction between MSSM and MSM if they are kinematically open. This requires that SUSY particles have masses already in the range accessible at LEP2. The total contribution of the other channels such as $h^0 \to \bar{c}c$, $\tau^+\tau^-$, gg, $\gamma\gamma$, $Z^0\gamma$ is less than 20% (10% for $\tan\beta \geq 3$). Deviations of the small branching fractions from the MSM expectations will be difficult to detect[3].

We assume that the MSSM and MSM can be distinguished experimentally if the difference between $BR(h^0 \to b\bar{b})$ is larger than about 8% [25]. Regions where this condition is fulfilled are shown in Fig. 11. The regions are defined as for the comparison of the cross sections. In the range about $m_A \geq 50$ GeV and $m_h \leq 80$ GeV, the knowledge of $BR(h^0 \to b\bar{b})$ at the level of 8% is insufficient to establish the nature of the scalar. In a large region (B), the perspectives of distinction depend on the SUSY parameters.

For $m_A \geq 100$ GeV the branching ratios in the MSSM and MSM differ for most choices of the SUSY parameters about 6%. This means that a very high experimental precision is required to identify the Higgs boson as a MSSM one.

It is important to note that although the above comparison is independent (branching ratios) or weakly dependent (cross sections) on the center-of-mass energy, the experimental feasibility depends strongly on the machine energy. This is due to the different numbers of expected Higgs events which are relevant for the achievable statistical significance of the cross section and branching ratio measurements.

## 8. Conclusions

Aspects of searches for neutral supersymmetric Higgs bosons at present and future $e^+e^-$ colliders have been presented. Full 1-loop diagrammatic calculations of radiative corrections to the Higgs particle production and decay rates are applied. The dependence of the results on all important model parameters is investigated. In addition to the stop mass, which is the most important free parameter in previous studies, several other SUSY parameters have been varied independently. This variation changes significantly the results compared with the simpler EPA approach. We show that for fixed and heavy SUSY particle masses (and small left-right sfermion mass splitting) the results of the EPA can approximately be recovered in the FDC. Nevertheless, differences of the order of few GeV on the investigated Higgs mass bounds also exist in this case. Using detailed experimental results and performing a full scan over the MSSM parameters, the differences become large. For LEP1, FDC gives in comparison with EPA an additional unexcluded region for $m_h \approx 25 - 50$ GeV and $m_A \approx 45 - 80$ GeV. We propose our method of interpretation to be adopted by the four LEP experiments, using higher statistics available now.

---

[3]Recently, a possible distinction of the $h^0$ from the $H^0_{MSM}$ on the basis of the off-shell decay $h^0 \to WW^\star$ has been discussed [24]

For LEP2, detailed experimental simulations for non-minimal Higgs bosons are combined with the improved calculations. Significant effects of the variation of the SUSY parameters on the accessible mass parameter ranges are found. The possibilities of a discovery of the lightest supersymmetric scalar depend strongly on the achievable center-of-mass energy. For $m_A > 120$ GeV, $h^0$ with the mass in the range $m_h < (\sqrt{s} - 100)$ GeV could always be found. In the $m_h$ range from $(\sqrt{s} - 100)$ GeV up to the kinematical bound, perspectives of discovery depend on the specified set of the MSSM parameters.

At the NLC even at the most unfavourable parameter choice at least one MSSM neutral Higgs boson should be found or the MSSM is ruled out. The NLC has good chances to discover more than one Higgs particles, and most likely either one or all three MSSM neutral scalars could be observed. If only one CP-even scalar is found via the bremsstrahlung process $e^+e^- \to Z^0 h^0$, it will be difficult to assess its origin. In a large mass region such a scalar could be distinguished from the MSM Higgs boson on the basis of its fermionic decay branching ratios if the measurements achieve a precision better then about 6%.

## Acknowledgements

We wish to thank Stefan Pokorski and Piotr Chankowski for the numerous and very fruitful discussions and their comments on the manuscript. This work is supported by the IFIC postdoctoral fellowship (J.R.) and by the Polish Committee for Scientific Research. A.S. wishes to thank for the hospitality during his stay in Warsaw.

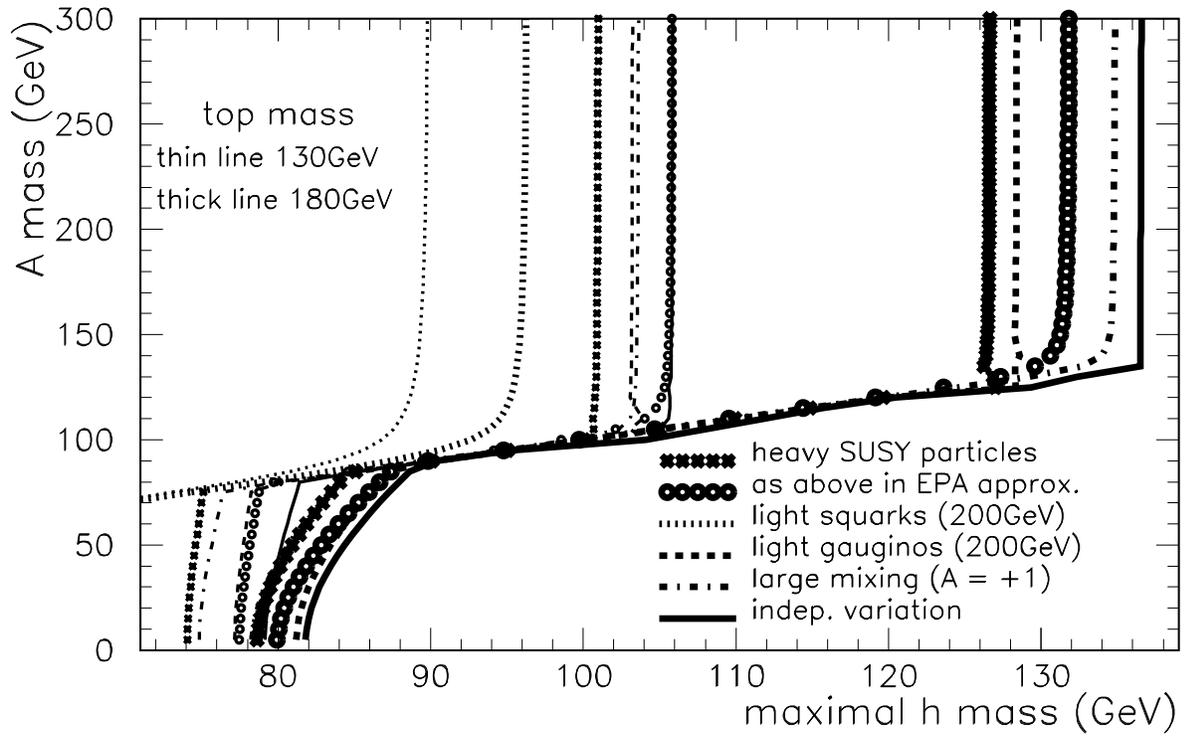

Figure 1: Upper bound for the h⁰ mass for various values of SUSY parameters.

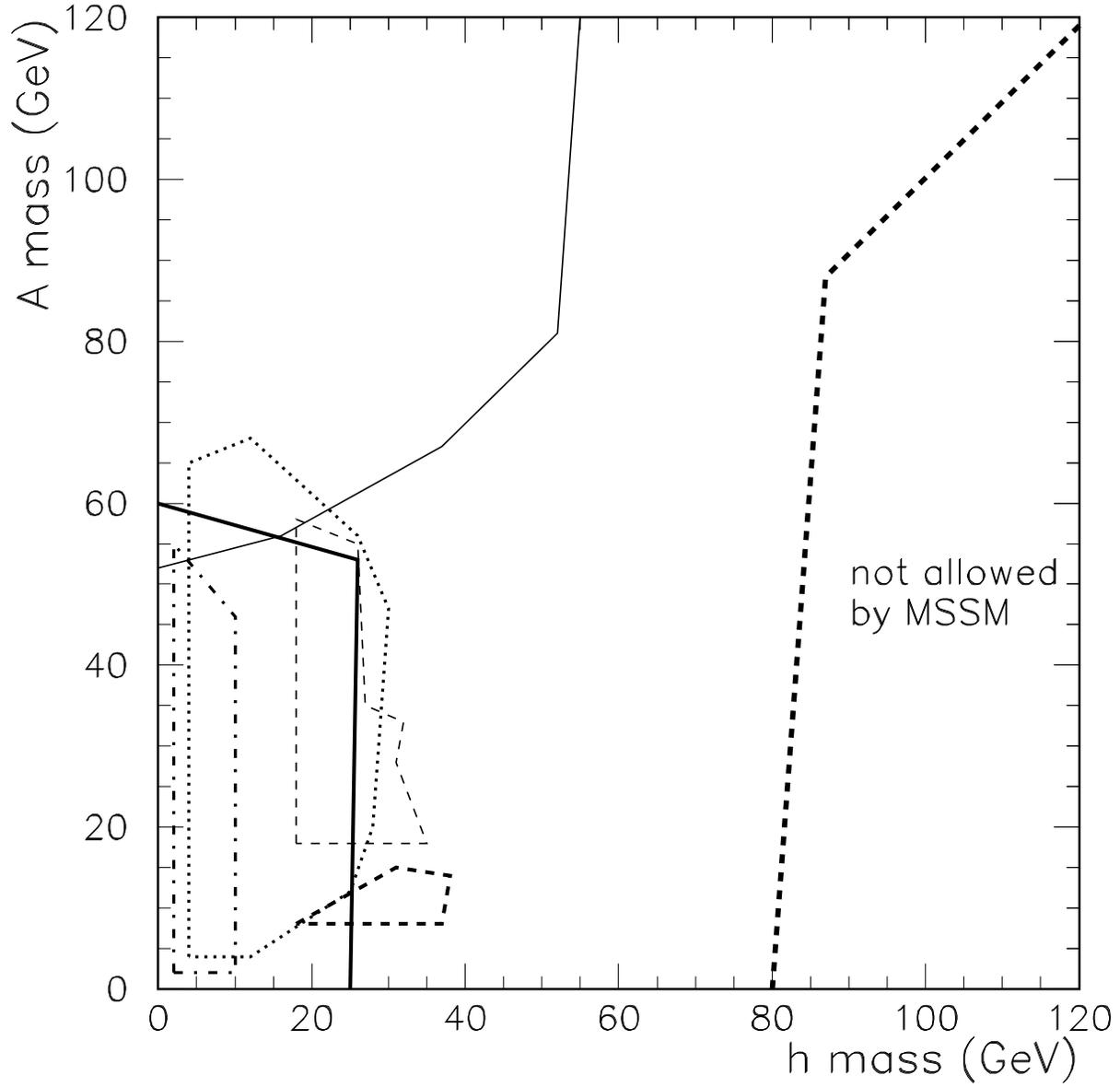

Figure 2: Individually excluded mass regions in the $(m_h, m_A)$ plane for $\sqrt{s} = m_Z$ for various searches: Thick solid line: $Z^0$ lineshape; thin solid line: $Z^0 \to Z^{0*}h^0$; dotted line: $\tau^+\tau^- b\bar{b}$; dotted-dashed line: $\tau^+\tau^-\tau^+\tau^-$; thin dashed: $b\bar{b}b\bar{b}$; thick dashed: 6b.

Figure 3: Excluded regions at $\sqrt{s} = m_Z$ in the $(m_h, m_A)$ plane. Dotted line: EPA (epsilon approximation). Thin solid line: FDC result for heavy SUSY parameters. Thick solid line: FDC results with a full scan over the SUSY parameters. Very thick line: new unexcluded region in FDC.

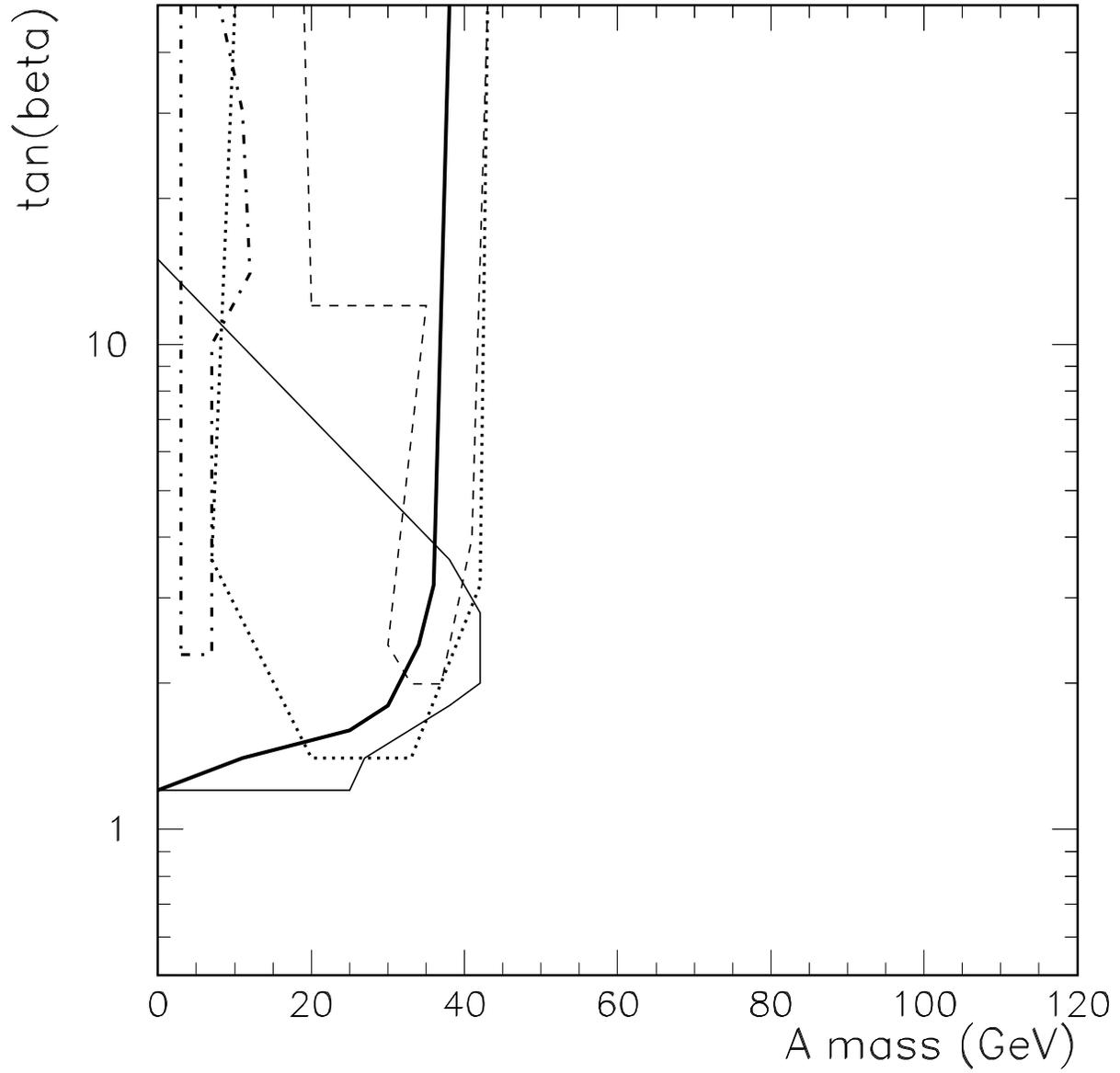

Figure 4: Individually excluded mass regions in the $(m_A, \tan\beta)$ plane for $\sqrt{s} = m_Z$ for various searches: Thick solid line: $Z^0$ lineshape; thin solid line: $Z^0 \to Z^{0*}h^0$; dotted line: $\tau^+\tau^- b\bar{b}$; dotted-dashed line: $\tau^+\tau^-\tau^+\tau^-$; thin dashed: $b\bar{b}b\bar{b}$.

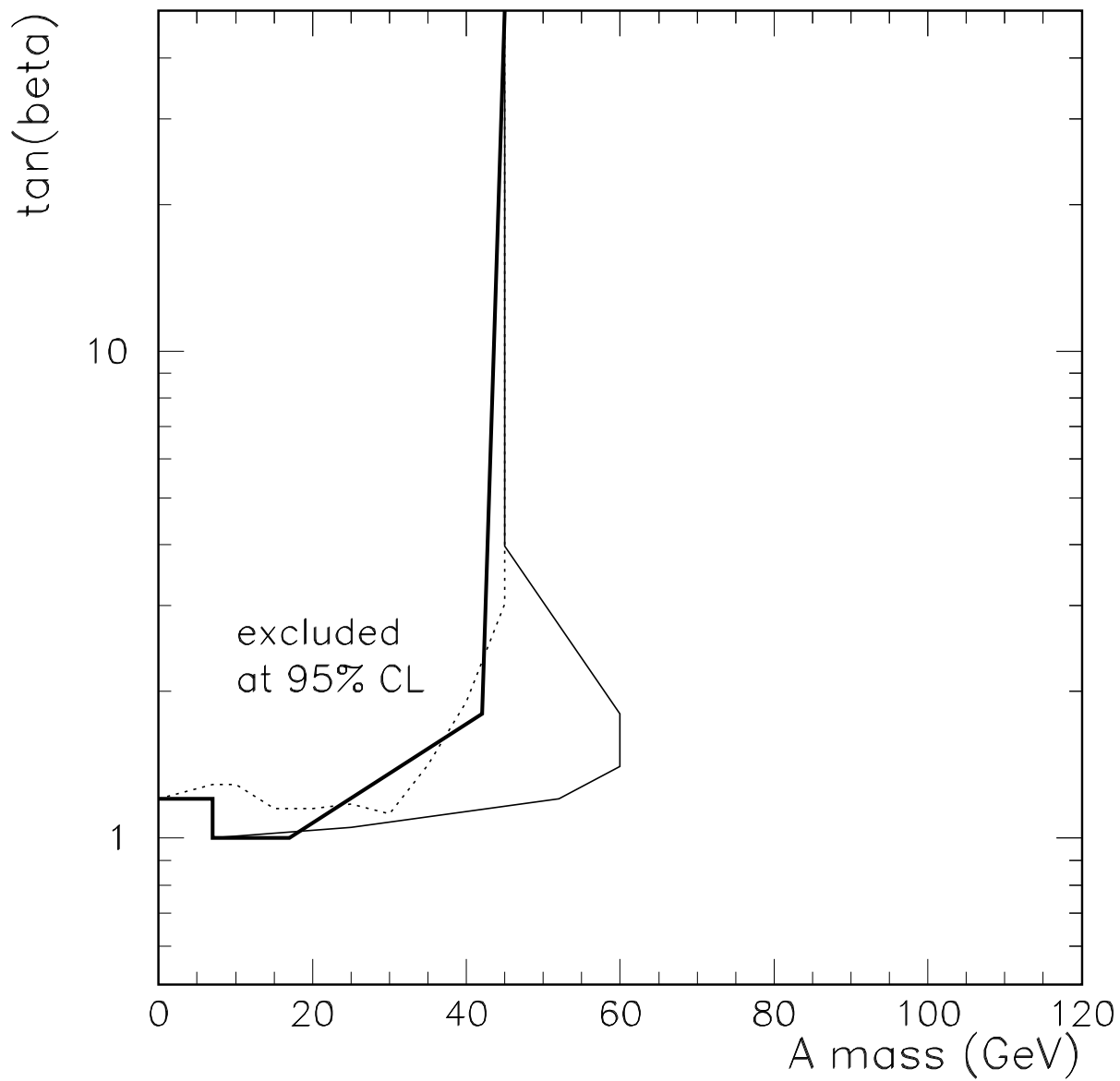

Figure 5: Excluded regions at $\sqrt{s} = m_Z$ in the $(m_A, \tan\beta)$. Dotted line: EPA (epsilon approximation). Thin solid line: FDC result for heavy SUSY parameters. Thick solid line: FDC results with a full scan over the SUSY parameters.

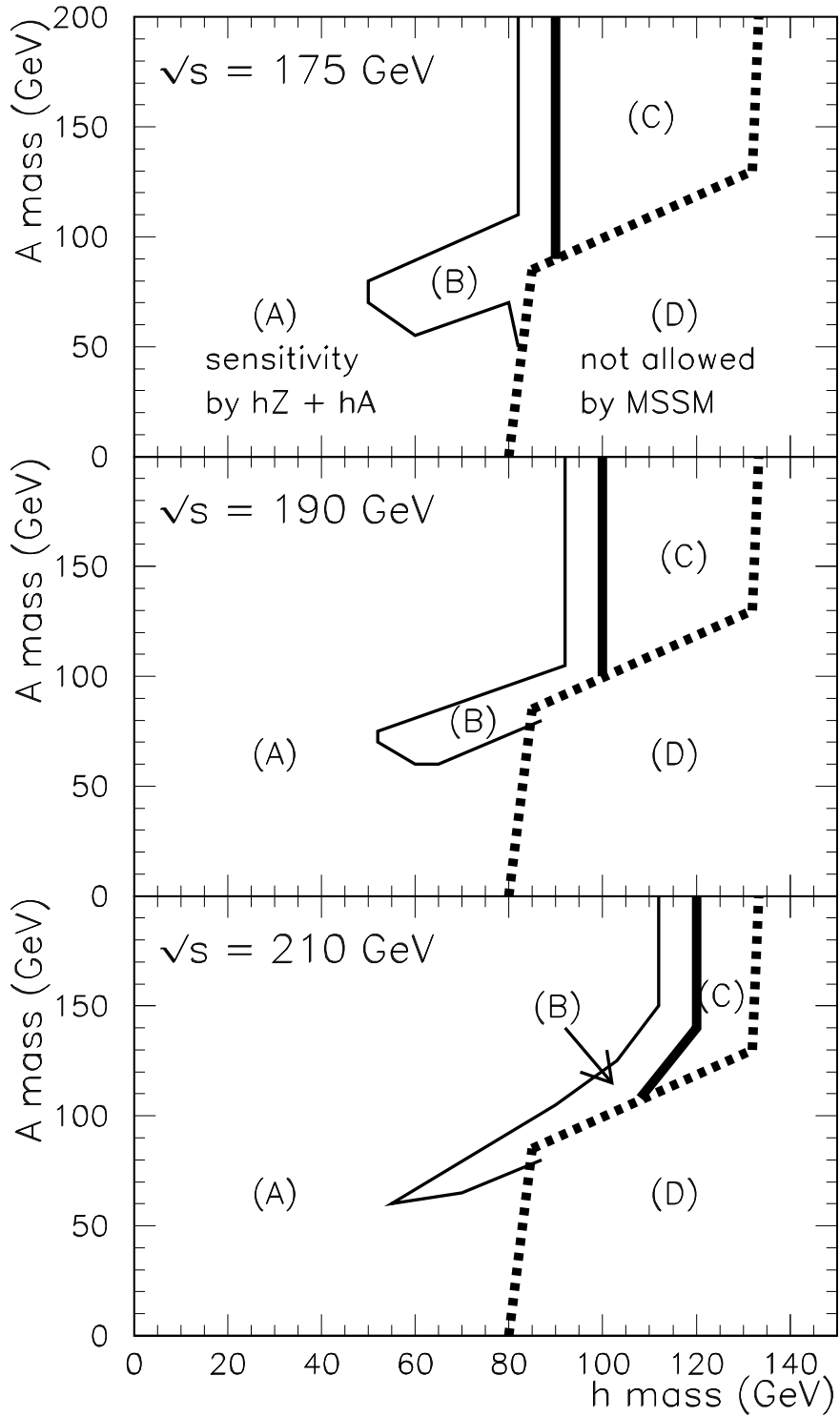

Figure 6: Regions of detectability of $h^0$ at LEP2 for $\sqrt{s}$ = 175, 190, and 210 GeV in the $(m_h, m_A)$ plane. For the description of regions (A)–(D) see text.

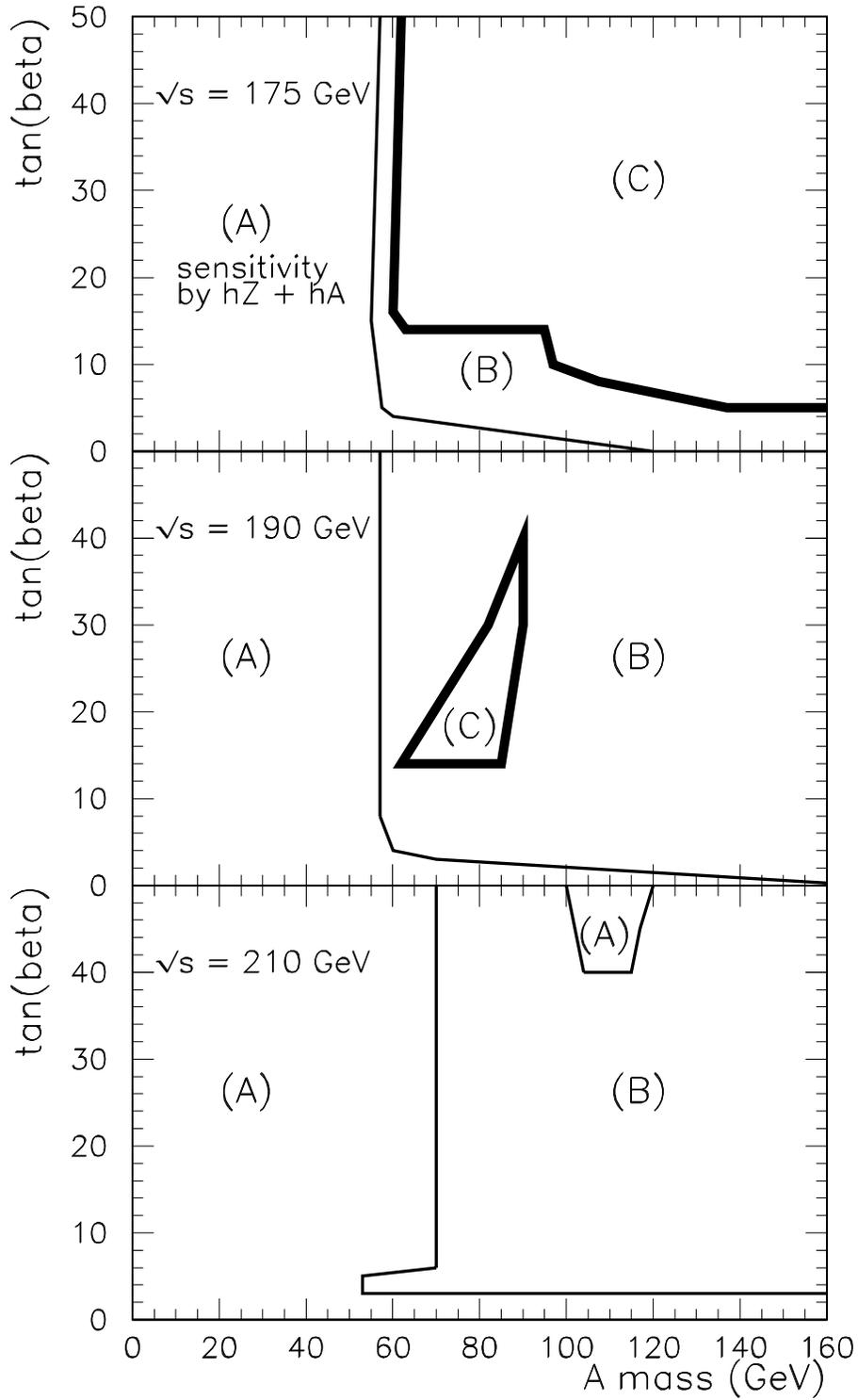

Figure 7: Regions of detectability of h⁰ at LEP2 for $\sqrt{s} = 175, 190,$ and $210$ GeV in the $(m_A, \tan\beta)$ plane. For the description of regions (A)–(C) see text.

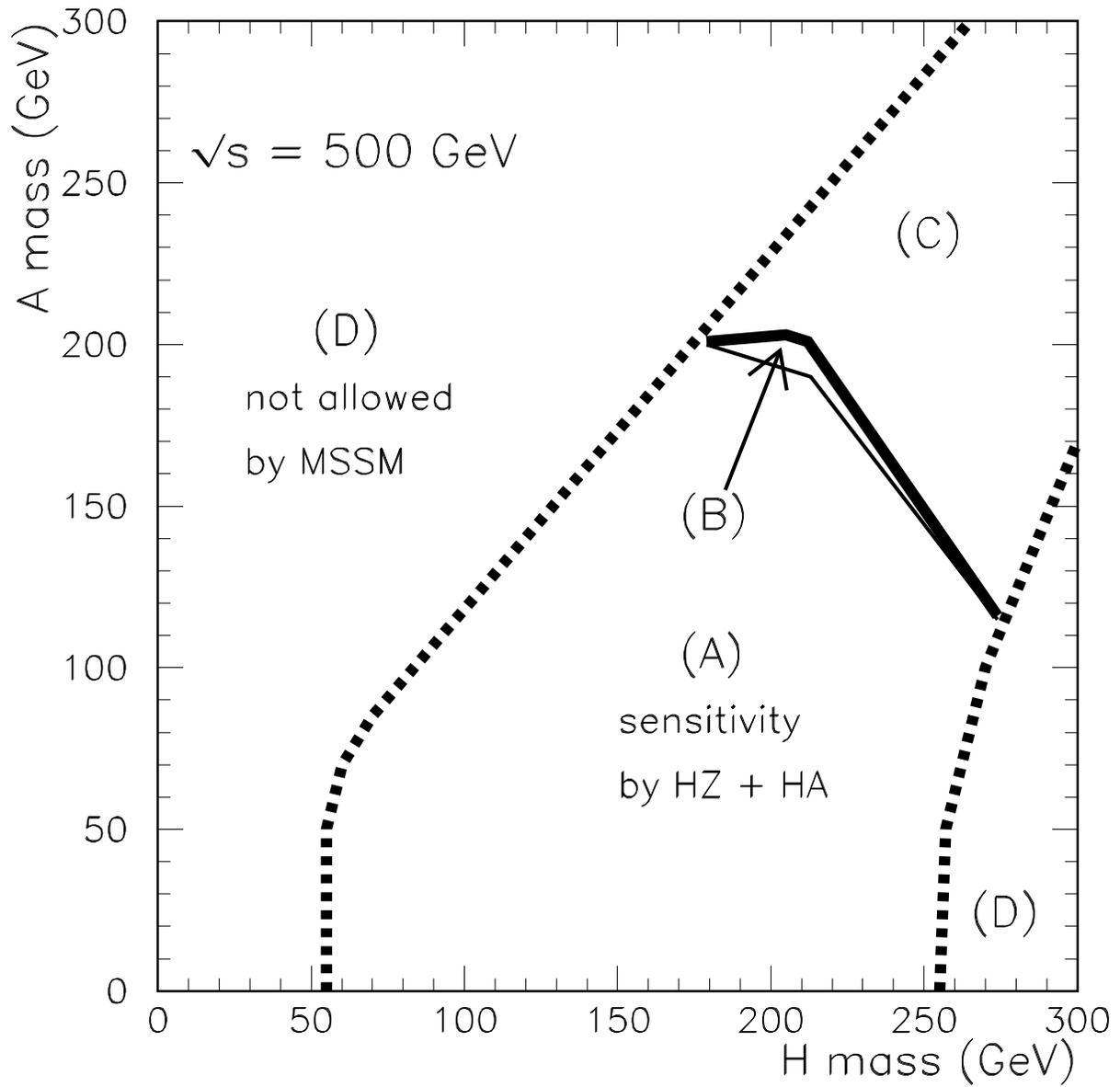

Figure 8: Regions of detectability of $H^0$ at the NLC for $\sqrt{s} = 500$ GeV. For the description of regions (A)–(D) see text.

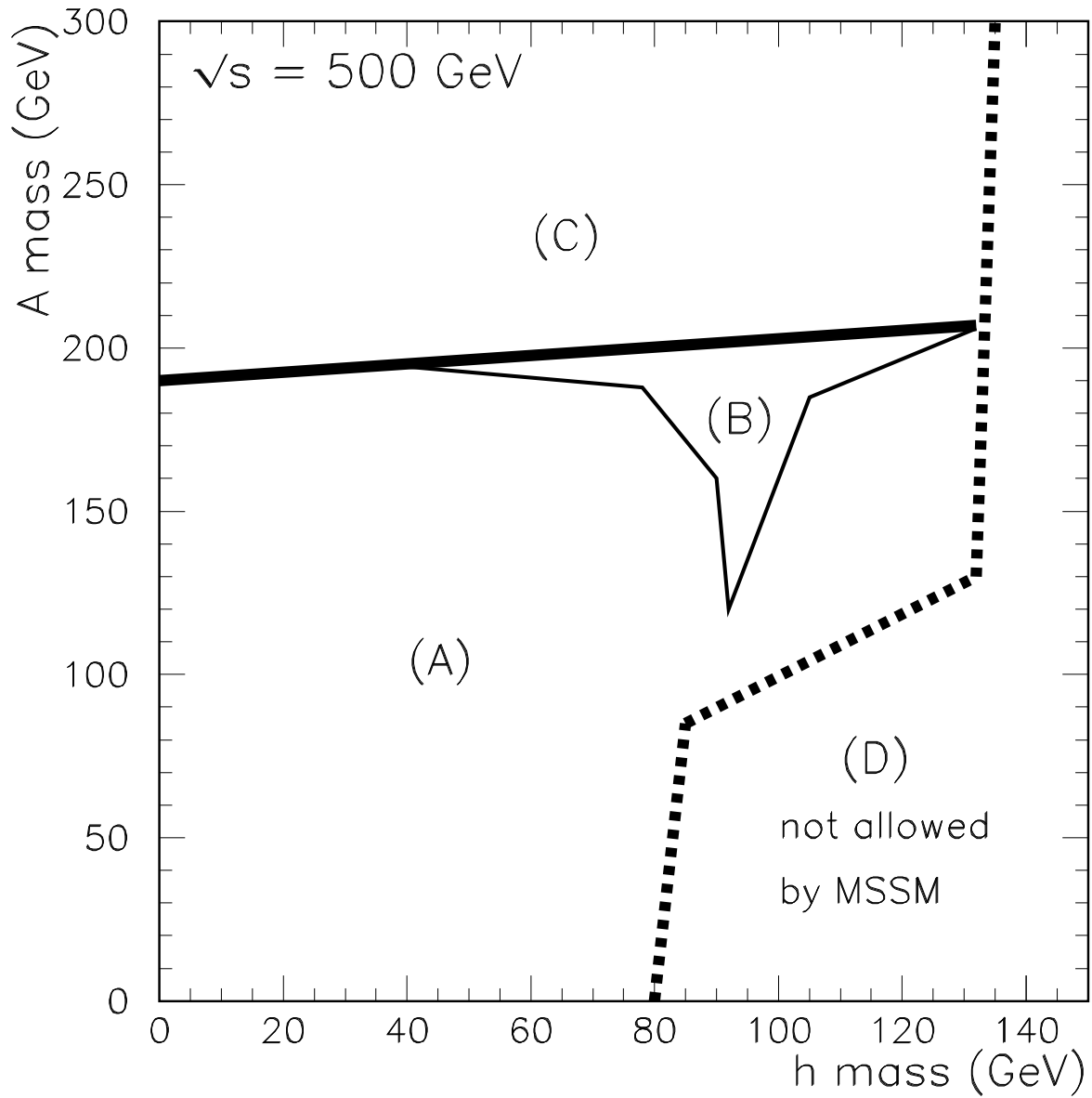

Figure 9: Regions of simultaneous detectability of $h^0$, $H^0$ and $A^0$ at the NLC for $\sqrt{s} = 500$ GeV. For the description of regions (A)–(D) see text.

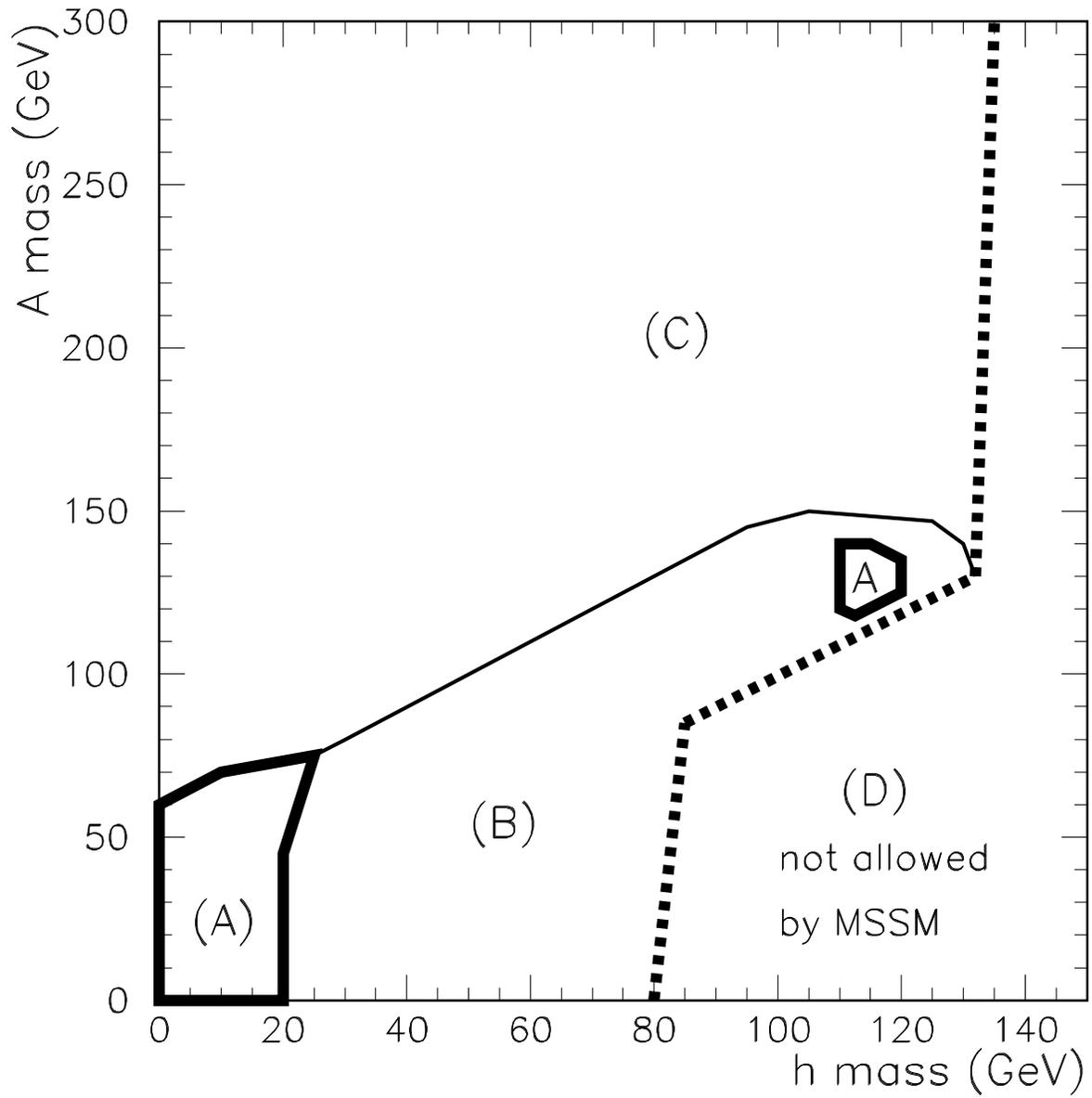

Figure 10: Regions where the MSSM and MSM Higgs boson can be distinguished on the basis of cross section differences for $e^+e^- \to Z^0 h^0$ reaction. Regions (A)–(C) show where the difference is larger (A) or smaller (C) than 20%, and where the results depend on the SUSY parameters (B).

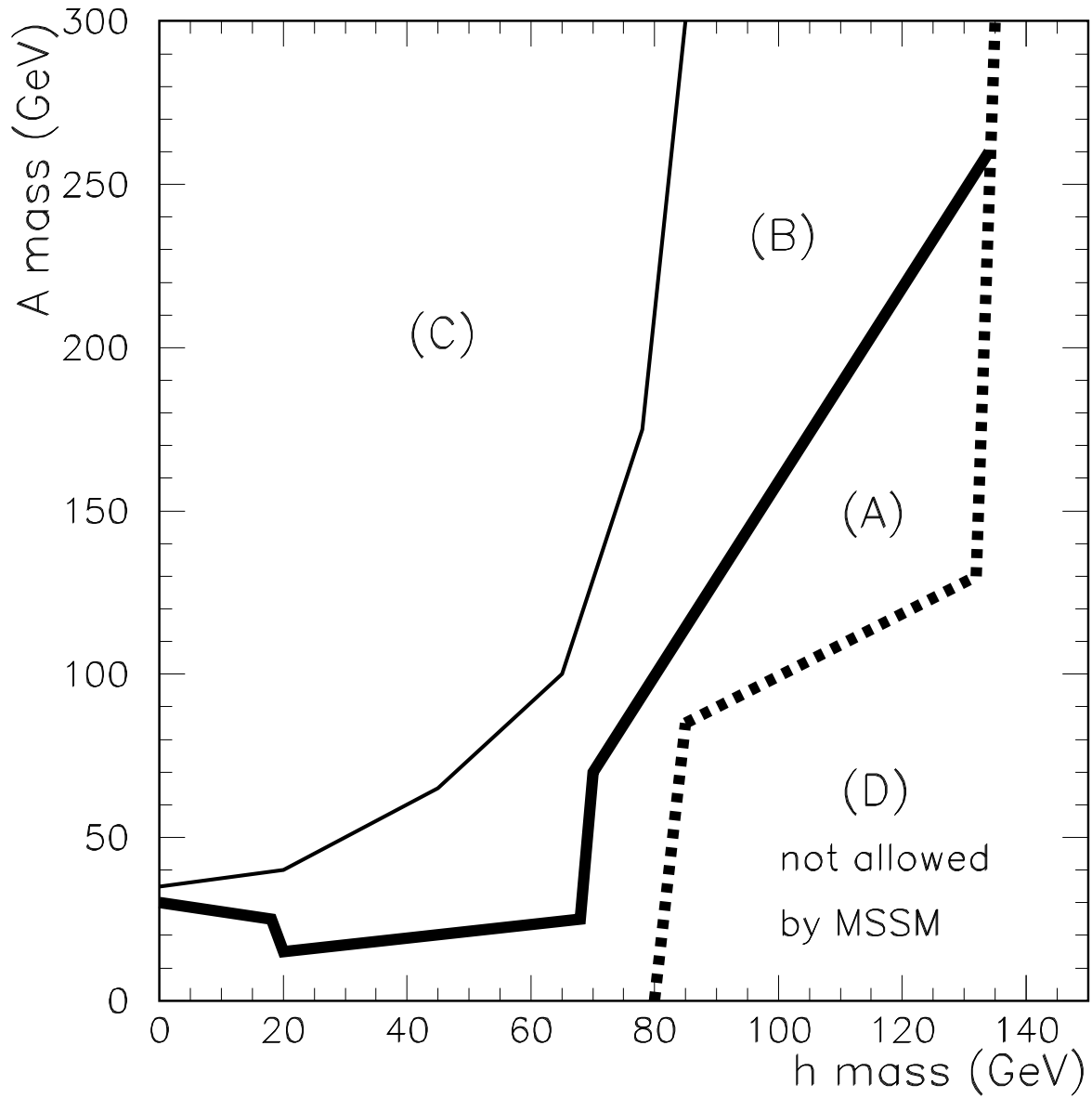

Figure 11: Regions where the MSSM and MSM Higgs boson can be distinguished on the basis of branching ratio differences for $\mathrm{BR}(h^0 \to b\bar{b})$ decay. Regions (A)–(C) show where the difference is larger (A) or smaller (C) than 8%, and where the results depend on the SUSY parameters (B).

24